\documentclass[12pt]{article}
\usepackage{graphicx}
\usepackage{lscape}
\usepackage{citesort}
\usepackage{amssymb}
\usepackage{appendix}
\usepackage{multirow}

\begin{document}
\def\qq{\langle \bar q q \rangle}
\def\uu{\langle \bar u u \rangle}
\def\dd{\langle \bar d d \rangle}
\def\sp{\langle \bar s s \rangle}
\def\GG{\langle g_s^2 G^2 \rangle}
\def\Tr{\mbox{Tr}}
\def\figt#1#2#3{
        \begin{figure}
        $\left. \right.$
        \vspace*{-2cm}
        \begin{center}
        \includegraphics[width=10cm]{#1}
        \end{center}
        \vspace*{-0.2cm}
        \caption{#3}
        \label{#2}
        \end{figure}
	}
	
\def\figb#1#2#3{
        \begin{figure}
        $\left. \right.$
        \vspace*{-1cm}
        \begin{center}
        \includegraphics[width=10cm]{#1}
        \end{center}
        \vspace*{-0.2cm}
        \caption{#3}
        \label{#2}
        \end{figure}
                }

\def\ds{\displaystyle}
\def\beq{\begin{equation}}
\def\eeq{\end{equation}}
\def\bea{\begin{eqnarray}}
\def\eea{\end{eqnarray}}
\def\beeq{\begin{eqnarray}}
\def\eeeq{\end{eqnarray}}
\def\ve{\vert}
\def\vel{\left|}
\def\ver{\right|}
\def\nnb{\nonumber}
\def\ga{\left(}
\def\dr{\right)}
\def\aga{\left\{}
\def\adr{\right\}}
\def\lla{\left<}
\def\rra{\right>}
\def\rar{\rightarrow}
\def\lrar{\leftrightarrow}  
\def\nnb{\nonumber}
\def\la{\langle}
\def\ra{\rangle}
\def\ba{\begin{array}}
\def\ea{\end{array}}
\def\tr{\mbox{Tr}}
\def\ssp{{\Sigma^{*+}}}
\def\sso{{\Sigma^{*0}}}
\def\ssm{{\Sigma^{*-}}}
\def\xis0{{\Xi^{*0}}}
\def\xism{{\Xi^{*-}}}
\def\qs{\la \bar s s \ra}
\def\qu{\la \bar u u \ra}
\def\qd{\la \bar d d \ra}
\def\qq{\la \bar q q \ra}
\def\gGgG{\la g^2 G^2 \ra}
\def\q{\gamma_5 \not\!q}
\def\x{\gamma_5 \not\!x}
\def\g5{\gamma_5}
\def\sb{S_Q^{cf}}
\def\sd{S_d^{be}}
\def\su{S_u^{ad}}
\def\sbp{{S}_Q^{'cf}}
\def\sdp{{S}_d^{'be}}
\def\sup{{S}_u^{'ad}}
\def\ssp{{S}_s^{'??}}

\def\sig{\sigma_{\mu \nu} \gamma_5 p^\mu q^\nu}
\def\fo{f_0(\frac{s_0}{M^2})}
\def\ffi{f_1(\frac{s_0}{M^2})}
\def\fii{f_2(\frac{s_0}{M^2})}
\def\O{{\cal O}}
\def\sl{{\Sigma^0 \Lambda}}
\def\es{\!\!\! &=& \!\!\!}
\def\ap{\!\!\! &\approx& \!\!\!}
\def\ar{&+& \!\!\!}
\def\ek{&-& \!\!\!}
\def\kek{\!\!\!&-& \!\!\!}
\def\cp{&\times& \!\!\!}
\def\se{\!\!\! &\simeq& \!\!\!}
\def\eqv{&\equiv& \!\!\!}
\def\kpm{&\pm& \!\!\!}
\def\kmp{&\mp& \!\!\!}
\def\mcdot{\!\cdot\!}
\def\erar{&\rightarrow&}


\def\simlt{\stackrel{<}{{}_\sim}}
\def\simgt{\stackrel{>}{{}_\sim}}


\renewcommand{\textfraction}{0.2}    
\renewcommand{\topfraction}{0.8}   

\renewcommand{\bottomfraction}{0.4}   
\renewcommand{\floatpagefraction}{0.8}
\newcommand\mysection{\setcounter{equation}{0}\section}

\def\baeq{\begin{appeq}}     \def\eaeq{\end{appeq}}  
\def\baeeq{\begin{appeeq}}   \def\eaeeq{\end{appeeq}}
\newenvironment{appeq}{\beq}{\eeq}   
\newenvironment{appeeq}{\beeq}{\eeeq}
\def\bAPP#1#2{
 \markright{APPENDIX #1}
 \addcontentsline{toc}{section}{Appendix #1: #2}
 \medskip
 \medskip
 \begin{center}      {\bf\LARGE Appendix #1 :}{\quad\Large\bf #2}
\end{center}
 \renewcommand{\thesection}{#1.\arabic{section}}
\setcounter{equation}{0}
        \renewcommand{\thehran}{#1.\arabic{hran}}
\renewenvironment{appeq}
  {  \renewcommand{\theequation}{#1.\arabic{equation}}
     \beq
  }{\eeq}
\renewenvironment{appeeq}
  {  \renewcommand{\theequation}{#1.\arabic{equation}}
     \beeq
  }{\eeeq}
\nopagebreak \noindent}

\def\eAPP{\renewcommand{\thehran}{\thesection.\arabic{hran}}}

\renewcommand{\theequation}{\arabic{equation}}
\newcounter{hran}
\renewcommand{\thehran}{\thesection.\arabic{hran}}

\def\bmini{\setcounter{hran}{\value{equation}}
\refstepcounter{hran}\setcounter{equation}{0}
\renewcommand{\theequation}{\thehran\alph{equation}}\begin{eqnarray}}
\def\bminiG#1{\setcounter{hran}{\value{equation}}
\refstepcounter{hran}\setcounter{equation}{-1}
\renewcommand{\theequation}{\thehran\alph{equation}}
\refstepcounter{equation}\label{#1}\begin{eqnarray}}


\newskip\humongous \humongous=0pt plus 1000pt minus 1000pt
\def\caja{\mathsurround=0pt}


\title{
         {\Large
                 {\bf
Strong coupling constants of light pseudoscalar mesons with heavy baryons
in QCD 
                 }
         }
      }

\author{\vspace{1cm}\\
{\small T. M. Aliev \thanks {e-mail:
taliev@metu.edu.tr}~\footnote{permanent address:Institute of
Physics,Baku,Azerbaijan}\,\,, K. Azizi \thanks {e-mail:
kazizi@dogus.edu.tr}\,\,, M. Savc{\i} \thanks
{e-mail: savci@metu.edu.tr}} \\
{\small Physics Department, Middle East Technical University,
06531 Ankara, Turkey }\\
{\small$^\ddag$ Physics Division,  Faculty of Arts and Sciences,
Do\u gu\c s University,} \\
{\small Ac{\i}badem-Kad{\i}k\"oy,  34722 Istanbul, Turkey}}

\date{}

\begin{titlepage}
\maketitle
\thispagestyle{empty}

\begin{abstract}
We calculate the  strong coupling constants of light pseudoscalar mesons with heavy 
baryons within the light cone QCD sum rules method. It is shown that
sextet--sextet, sextet--antitriplet and antitriplet--antitriplet transitions
 are described by one universal invariant 
function for each class. A comparison of our results on the coupling constants 
with the predictions existing in literature is also presented. 
\end{abstract}

~~~PACS number(s): 11.55.Hx, 13.75.Gx, 13.75.Jz
\end{titlepage}

\section{Introduction}

In this decade exciting experimental results have been obtained in heavy
baryon spectroscopy. During these years, the ${1\over 2}^+$ and 
${1\over 2}^-$ antitriplet states, $\Lambda_c^+,~\Xi_c^+,~\Xi_c^0$ 
and $\Lambda_c^+ (2593)$,\\
$\Xi_c^+(2790),~\Xi_c^0(2790)$ and the ${1\over 2}^+$
and  ${3\over 2}^+$ and sextet states,
$\Omega_c^\ast,\Sigma_c^\ast,\Xi_c^\ast$ have been observed in experiments
\cite{Rstp01}. Among the s--wave bottom hadrons, only
$\Lambda_b,~\Sigma_b,~\Sigma_b^\ast,~\Xi_b$ and $\Omega_b$ have been discovered. 
Moreover, in recent years many new states have been observed by BaBar and
BELLE collaborations, such as,
$X(3872),~Y(3930),~Z(3930),~X(3940),~Y(4008),~Z_1^+(4050),$\\
$Y(4140),~X(4160),~Z_2(4250),~Y(4260),~Y(4360),~Z^+(4430),$
and $Y(4660)$ which remain unidentified.

Of course, establishing these states is a remarkable progress in hadron
physics. It is expected that LHC, the world's largest--highest--energy particle
accelerator, will open new horizons in the discovery of the excited bottom 
baryon sates \cite{Rstp02}. The experimental progress on heavy hadron
spectroscopy stimulated intensive theoretical studies in this respect (for a
review see \cite{Rstp03,Rstp04} and references therein). A detailed theoretical
study of experimental results on hadron spectroscopy and various weak and
strong decays can provide us with useful information about the quark
structure of new hadrons at the hadronic scale.

This scale belongs to the nonperturbative sector of QCD. Therefore, for
calculation of the form factors in weak decays and coupling constants in
strong decays, some nonperturbative methods are needed. Among many nonperturbative
methods, QCD sum rules \cite{Rstp05} is more reliable and predictive. In the
present work, we calculate the strong coupling constants of light
pseudoscalar mesons  with sextet and antitriplet baryons, in light cone version
of the QCD sum rules (LCSR) method (for a review, see \cite{Rstp06}).
Note that some of the strong coupling constants have already been studied in
\cite{Rstp07,Rstp08,Rstp09} in the same framework.

The outline of this article is as follows. In section 2, we demonstrate how
coupling constants of pseudoscalar mesons with heavy baryons can be
calculated. In this section, the LCSR for the heavy baryon--pseudoscalar meson
coupling constants are also derived using the most general form of the baryon
currents. Section 3 is devoted to the numerical analysis and a comparison of
our results with the existing predictions in the literature.

\section{Light cone QCD sum rules for the coupling constants of
pseudoscalar mesons with heavy baryons}

Before presenting the detailed calculations for the strong coupling
constants of pseudoscalar mesons with heavy baryons, we would like to make few
remarks about the classification of heavy baryons. Heavy baryons with a
single heavy quark belong to either $SU(3)$ antisymmetric $\bar{3}_F$ or
symmetric $6_F$ flavor representations. Since we consider the ground states, the total spin of the two light quarks must one for $6_F$
and zero for $\bar{3}_F$, due to the symmetry property of their colors and
flavors, as a result of which we can write $J^P={1\over 2}^+ / {3\over 2}^+$
for $6_F$ and $J^P={1\over 2}^+$ for $\bar{3}_F$. Graphically, $6_F$ and
$\bar{3}_F$ representations are given in Fig. (1), where
$\alpha$, $\alpha +1$, $\alpha +2$ determine the charges of baryons
$(\alpha=-1$ or $ 0)$, and the asterix $(\ast)$ denote $J^P={3\over 2}^+$ 
states. In this work, we will consider only $J^P={1\over 2}^+$ states.

After this preliminary remarks, we proceed by calculating the strong
coupling constants of pseudoscalar mesons with heavy baryons within the
LCSR. For this purpose, we start by considering the following correlation
function:
\bea
\label{estp01}
\Pi^{(ij)} = i \int d^4x e^{ipx} \lla {\cal P}(q) \vel {\cal T} \left\{
\eta^{(i)} (x) \bar{\eta}^{(j)} (0) \right\} \ver 0 \rra~,
\eea
where ${\cal P}(q)$ is the pseudoscalar--meson  with 
momentum $q$, $\eta$ is the interpolating current for the heavy baryons and
${\cal T}$ is the time ordering operator. Here, $i=1,~j=1$ describes the
sextet--sextet, $i=1,~j=2$ corresponds to sextet--triplet, and  $i=2,~j=2$
describes triplet--triplet transitions. For convenience we shall denote
$\Pi^{(11)} = \Pi^{(1)}$, $\Pi^{(12)} = \Pi^{(2)}$ and $\Pi^{(22)} =
\Pi^{(3)}$.
The sum rules for the coupling constants of pseudoscalar mesons with heavy baryons
can be obtained by calculating the correlation function
(\ref{estp01}) in two different ways, namely, in terms of the hadrons and in
terms of quark gluon degrees of freedom, and then matching these two
representations.

Firstly, we calculate the correlation function (\ref{estp01}) in terms of
hadrons. Inserting  complete sets of hadrons with the same quantum numbers
in the interpolating currents and isolating the ground states, we obtain
\bea
\label{estp02}
\Pi^{(ij)} =  {\lla 0 \vel \eta^{(i)}(0) \ver
B_2(p) \rra \lla B_2(p) {\cal P}(q) \vel \right.  
B_1(p+q) \rra \lla B_1(p+q) \vel \bar{\eta}^{(j)}(0) \ver
0 \rra \over \ga p^2-m_2^2 \dr \left[(p+q)^2-m_1^2\right]}
+ \cdots~,
\eea  
where $\vel B_2(p) \rra$ and $\vel B_1(p+q) \rra$ are the ${1\over 2}$ states,
and $m_2$ and $m_1$ are their masses, respectively. The dots in
Eq. (\ref{estp02}) describe contributions of the higher states and continuum. It follows from Eq. (\ref{estp02}) that in order to calculate the
correlation function in terms of hadronic parameters, the matrix elements entering to
Eq. (\ref{estp02}) are needed. These matrix elements are defined
in the following way:
\bea
\label{estp03}
\lla 0 \vel \eta^{(i)} \ver B(p) \rra \es \lambda_i u(p) ~, \nnb \\
\lla B(p+q) \vel \eta^{(j)} \ver 0 \rra \es \lambda_j \bar{u}(p+q)~, \nnb \\
\lla B(p) {\cal P}(q) \vel \right. B(p+q) \rra \es g \bar{u}(p) i\gamma_5 
u(p+q)~,
\eea
where $\lambda_i$ and $\lambda_j$ are the residues of the heavy baryons, 
$g$ is the coupling constant of pseudoscalar meson with heavy baryon and $u$ 
is the Dirac bispinor.

Using Eqs. (\ref{estp02}) and (\ref{estp03}) and performing summation
over spins of the baryons, we obtain the following representation of the
correlation function from the hadronic side:
\bea
\label{estp04}
\Pi^{(ij)} \es i {\lambda_i \lambda_j g \over \ga p^2-m_2^2 \dr
\left[(p+q)^2-m_1^2\right]} \Big\{ \rlap/q\rlap/p  \gamma_5 + \mbox{other structures} \Big\},\nnb\\
\eea
where we kept the  structure which leads to a more  reliable result.

In order to calculate the correlation function from QCD side, the forms of
the interpolating currents for the heavy baryons are needed.  The general
form of the interpolating currents for the heavy spin ${1\over 2}$ sextet
and antitriplet baryons can be written as (see for example \cite{E.Bagan}),
\bea
\label{estp05}
\eta_Q^{(s)} \es - {1\over \sqrt{2}} \epsilon^{abc} \Big\{ \Big( q_1^{aT} 
C Q^b \Big) \gamma_5 q_2^c + \beta \Big( q_1^{aT} C \gamma_5 Q^b \Big) q_2^c -
\Big[\Big( Q^{aT} C q_2^b \Big) \gamma_5 q_1^c + \beta \Big( Q^{aT} C
\gamma_5 q_2^b \Big) q_1^c \Big] \Big\}~, \nnb\\
\label{estp06} 
\eta_Q^{(anti-t)} \es {1\over \sqrt{6}} \epsilon^{abc} \Big\{ 2 \Big( q_1^{aT} 
C q_2^b \Big) \gamma_5 Q^c + 2 \beta \Big( q_1^{aT} C \gamma_5 q_2^b \Big) Q^c
+ \Big( q_1^{aT} C Q^b \Big) \gamma_5 q_2^c + \beta \Big(q_1^{aT} C
\gamma_5 Q^b \Big) q_2^c \nnb \\
\ar \Big(Q^{aT} C q_2^b \Big) \gamma_5 q_1^c +
\beta \Big(Q^{aT} C \gamma_5 q_2^b \Big) q_1^c \Big\}~,
\eea
where $a,,b,c$ are the color indices and  $\beta$ is an arbitrary parameter. It should also  be noted that the general form of interpolating currents for light
 spin 1/2 baryons was introduced in \cite{Y.Chang} and 
$\beta=-1$ corresponds to the Ioffe current \cite{B.L.Ioffe}. The quark fields $q_1$ and
$q_2$ for the sextet and antitriplet are presented in Table 1.


\begin{table}[h]

\renewcommand{\arraystretch}{1.3}
\addtolength{\arraycolsep}{-0.5pt}
\small
$$
\begin{array}{|l|c|c|c|c|c|c|c|c|c|}
\hline \hline 
 & \Sigma_{b(c)}^{+(++)} &\Sigma_{b(c)}^{0(+)} &\Sigma_{b(c)}^{-(0)} & \Xi_{b(c)}^{-(0)'} &\Xi_{b(c)}^{0(+)'}  &\Omega_{b(c)}^{-(0)} &\Lambda_{b(c)}^{0(+)} &
 \Xi_{b(c)}^{-(0)}& \Xi_{b(c)}^{0(+)} \\  \hline
q_1&u &u & d&d  &u &s &u&d &u \\
q_2&u & d& d&s & s&s &d &s &s\\
\hline \hline
\end{array}
$$
\caption{The quark flavors $q_1$ and $q_2$ for the baryons in the sextet and
the antitriplet representations}
\renewcommand{\arraystretch}{1}
\addtolength{\arraycolsep}{-1.0pt}

\end{table}
 
As has already been noted, in order to calculate the coupling constants of
pseudoscalar mesons with heavy baryons entering to sextet and
antitriplet representation, the calculation of the correlation function
from QCD part is needed. Before calculating it, we follow the approach given
in \cite{Rstp10,Rstp11,Rstp12,Rstp13,Rstp14} and try to find relations among
invariant functions involving coupling constants of pseudoscalar mesons with
sextet and antitriplet baryons.
We will show that the correlation functions responsible for coupling
constants of pseudoscalar mesons (P) with sextet--sextet (SS),
sextet--antitriplet (SA)  and antitriplet-antitriplet (AA) baryons can each
be represented in terms of only one invariant function. Of course, the form
of the invariant functions for the couplings SSP, SAP and AAP are different
in the general case. It should be noted here that  the relations presented 
below are all structure independent.

We start our discussion by considering the sextet--sextet transition,
concretely. Consider the $\Sigma_b^0 \rar \Sigma_b^0 \pi^0$ transition. The
invariant function for this transformation can be written in the following
form
\bea
\label{estp07}  
\Pi^{\Sigma_b^0 \rar \Sigma_b^0 \pi^0} = g_{\pi\bar{u}u} \Pi_1^{(1)}(u,d,b) +
g_{\pi\bar{d}d} \Pi_1^{'(1)}(u,d,b) + g_{\pi\bar{b}b} \Pi_2^{(1)}(u,d,b)~,
\eea
where the interpolating current of $\pi^0$ meson is written as
\bea
\label{nolabel} 
J_{\pi^0} = \sum_{u,d} g_{\pi \bar{q}q}\bar{q}\gamma_5 q~. \nnb
\eea

Obviously, the relations $g_{\pi \bar{u}u} = -g_{\pi \bar{d}d}={1\over \sqrt{2}}$, $g_{\pi
\bar{b}b} = 0$ hold for the $\pi^0$ meson. The invariant functions
$\Pi_1,~\Pi_1^{'}$ and $\Pi_2$ describe the radiation of $\pi^0$
meson from $u,~d$ and $b$ quarks of $\Sigma_b^0$ baryon, respectively, and 
they can formally be defined as:
\bea
\label{estp08}
\Pi_1^{(1)}(u,d,b) \es \lla \bar{u}u \vel \Sigma_b^0 \bar{\Sigma}_b^0 \ver 0
\rra~, \nnb \\
\Pi_1^{'(1)}(u,d,b) \es \lla \bar{d}d \vel \Sigma_b^0 \bar{\Sigma}_b^0 \ver 0
\rra~, \nnb \\
\Pi_2^{(1)}(u,d,b) \es \lla \bar{b}b \vel \Sigma_b^0 \bar{\Sigma}_b^0 \ver 0
\rra~.
\eea
It follows from the definition of the interpolating current of $\Sigma_b$
baryon that it is symmetric under the exchange $u \lrar d$, hence
$\Pi_1^{'(1)}(u,d,b) = \Pi_1^{(1)}(d,u,b)$. Using this relation, we 
immediately get from Eq. (\ref{estp07}) that,
\bea
\label{estp09}
\Pi^{\Sigma_b^0 \rar \Sigma_b^0 \pi^0} = {1\over \sqrt{2}} \Big[
\Pi_1^{(1)}(u,d,b) - \Pi_1^{(1)}(d,u,b) \Big]~,
\eea
and one can easily see that in the $SU_2(2)_f$ limit, $\Pi^{\Sigma_b^0 \rar
\Sigma_b^0 \pi^0} = 0$.

The invariant function responsible for the $\Sigma_b^+ \rar \Sigma_b^+
\pi^0$ transition can be obtained from the $\Sigma_b^0 \rar \Sigma_b^0
\pi^0$ case by making the replacement $d \rar u$, and using
$\Sigma_b^0=-\sqrt{2} \Sigma_b^+$, from which we get,
\bea            
\label{estp10}
4 \Pi_1^{(1)}(u,d,b) = -2 \lla \bar{u}u \vel \Sigma_b^+ \bar{\Sigma}_b^+ \ver 0
\rra~.
\eea
Appearance of the factor 4 on the left hand side is due to the fact that
each $\Sigma^+_b$ contains two $u$ quark, hence there are 4 possible ways for
radiating $\pi^0$ from the $u$ quark.
Making use of Eq.(\ref{estp09}), we get
\bea            
\label{estp11}
\Pi^{\Sigma_b^+ \rar \Sigma_b^+ \pi^0} = \sqrt{2} \Pi_1^{(1)}(u,u,b).
\eea

The invariant function describing $\Sigma_b^- \rar \Sigma_b^- \pi^0$ can
easily be obtained from the $\Sigma_b^0 \rar \Sigma_b^0 \pi^0$ transition by
making the replacement $u \rar d$ and taking into account $\Sigma_b^0 (u\rar
d) = \sqrt{2} \Sigma_b^-$. Performing calculation similar to the previous
case, we get
\bea            
\label{estp12}
\Pi^{\Sigma_b^- \rar \Sigma_b^- \pi^0} = \sqrt{2} \Pi_1^{(1)}(d,d,b).
\eea

Now, let us proceed to obtain the results for the invariant function
involving $\Xi_b^{'-(0)} \rar \Xi_b^{'-(0)} \pi^0$ transition. The invariant
function for this transition can be obtained from the  $\Sigma_b^0 \rar
\Sigma_b^0 \pi^0$ case using the fact that $\Xi_b^{'0} = \Sigma_b^0
(d\rar s)$ and  $\Xi_b^{'-} = \Sigma_b^0
(u\rar s)$. As a result, we obtain
\bea
\label{estp13}
\Pi^{\Xi_b^{'0} \rar \Xi_b^{'0} \pi^0} \es {1\over \sqrt{2}}
\Pi_1^{(1)}(u,s,b)~, \nnb \\
\Pi^{\Xi_b^{'-} \rar \Xi_b^{'-} \pi^0} \es - {1\over \sqrt{2}}
\Pi_1^{(1)}(d,s,b)~.
\eea

Obtaining relations among the invariant functions involving charged $\pi^\pm$
mesons requires more care. In this respect, we start by considering the
matrix element $\lla \bar{d}d \vel\Sigma_b^0 \bar{\Sigma}_b^0 \ver 0 \rra$,
where $d$ quarks from the $\Sigma_b^0$ and $\bar{\Sigma}_b^0$ from the
final $\bar{d}d$ state, and $u$ and $b$ quarks are the spectators. The
matrix element $\lla \bar{u}d \vel\Sigma_b^+ \bar{\Sigma}_b^0 \ver 0
\rra$describes the case where $d$ quark from $\bar{\Sigma}_b^0$ and $u$ 
quark from $\Sigma_b^+$ form the $\bar{u}d$ state and the remaining $u$ and
$b$ are being again the spectators. One can expect from this observation
that  these matrix elements should be proportional to each other and
calculations confirm this expectation. So,
\bea
\label{estp16}
\Pi^{\Sigma_b^0 \rar \Sigma_b^+ \pi^-} \es \lla \bar{u}d \vel \Sigma_b^+
\bar{\Sigma}_b^0 \ver 0 \rra = -\sqrt{2} \lla \bar{d}d \vel \Sigma_b^0
\bar{\Sigma}_b^0 \ver 0 \rra = -\sqrt{2} \Pi_1^{(1)}(d,u,b)~.
\eea
Making the replacement $u \lrar d$ in Eq. (\ref{estp16}), we obtain
\bea
\label{estp17}  
\Pi^{\Sigma_b^0 \rar \Sigma_b^- \pi^+} \es \lla \bar{d}u \vel \Sigma_b^-
\bar{\Sigma}_b^0 \ver 0 \rra = \sqrt{2} \lla \bar{u}u \vel \Sigma_b^0
\bar{\Sigma}_b^0 \ver 0 \rra = \sqrt{2} \Pi_1^{(1)}(u,d,b)~.        
\eea

In estimating the coupling constants of SSP, SAP and AAP, it is enough to 
consider the $\Sigma_b^0 \rar \Sigma_b^0 P$, $\Xi_b^{'0} \rar \Xi_b^0 P$ and
$\Xi_b^0 \rar \Xi_b^0 P$ transitions, respectively. All remaining
transitions can be obtained from these transitions with the help of the
appropriate transformations among quark fields. Relations among the invariant functions of the charmed baryons can easily be obtained by making the 
the
replacement $b \rar c$ and adding to charge of each baryon a positive unit
charge. 

Performing similar calculations, one can obtain rest of the required 
expressions from the correlation functions in terms of the invariant 
function $\Pi_1$, involving $\pi$, $K$ and $\eta$ mesons describing 
sextet--sextet, sextet--antitriplet and antitriplet--antitriplet 
transitions. In the present work, we  neglect the mixing between $\eta$ and $\eta^{'}$
mesons.
It should also be noted here that all coupling constants for the SSP, SAP and AAP
are described only by one invariant function in each class of transitions, 
but the forms of the invariant functions in each group of transitions are
different. 

The invariant function $\Pi_1$ responsible for the $\Sigma_b^0 \rar
\Sigma_b^0 P$, $\Xi_b^{'0} \rar \Xi_b^0 P$ and $\Xi_b^0 \rar \Xi_b^0 P$ 
transitions can be calculated in
deep Euclidean region, $-p^2 \rar +\infty$ and $-(p+q)^2 \rar +\infty$ using the
operator product expansion (OPE) in terms of the distribution amplitudes (DA's) of
the pseudoscalar mesons and light and heavy quark operators. Up to twist--4
accuracy, the matrix elements $\lla P(q) \vel \bar{q}(x) \Gamma q(0) \ver 0
\rra$ and $\lla P(q) \vel \bar{q}(x) G_{\mu\nu} q(0) \ver 0 \rra$, where
$\Gamma$ is any arbitrary Dirac matrix, are determined in terms of the DA's 
of the pseudoscalar mesons, and their explicit expressions are given in 
\cite{Rstp15,Rstp16,Rstp17}.

The light and heavy quark propagators are  calculated in
\cite{Rstp18}, and
\cite{Rstp20}, respectively.
Using expressions of these  propagators and
definitions of DA's for the pseudoscalar mesons, the correlation function
can be calculated from the QCD side, straightforwardly.
Equating the coefficients of the structure $\rlap/q\rlap/p \gamma_5$ of the
representation of the correlation function from hadronic and theoretical
sides, and applying the Borel transformation with respect to the variables
$p^2$ and $(p+q)^2$ in order to suppress the contributions of the higher
states and continuum, we obtain the following sum rules for the strong
coupling constants of the pseudoscalar mesons with sextet and antitriplet
baryons:
\bea
\label{estp20}
g^{(i)} = {1 \over \lambda_1^{(i)} \lambda_2^{(i)}} e^{{m_1^{(i)2} 
\over M_1^2} + {m_2^{(i)2}
\over M_2^2} }\, \Pi_1^{(i)}~,
\eea
where, $i=1,~2$ and $3$ for  sextet--sextet,  sextet--antitriplet and antitriplet--antitriplet, respectively and $M_1^2$ and $M_2^2$ are the Borel masses corresponding to
the initial and the final baryons. Since the masses of the initial and final
baryons are practically equal to each other, we take $M_1^2=M_2^2=2 M^2$;
and $\lambda_1^{(i)}$ and $\lambda_2^{(i)}$ are the residues of the initial and
final baryons, respectively, which are calculated in \cite{Rstp21}.
The explicit expressions for $\Pi_1^{(i)}$ are quite lengthy and we do not
present all of them here.  As an example, we only present the explicit expression of the $\Pi_1^{(1)}$, which is given as:
\bea
&&e^{m_Q^2/M^2 - m_{\cal P}^2/M^2} \Pi_1^{(1)} (u,d,b) = \nnb \\ 
&&{(1-\beta)^2\over 32 \sqrt{2} \pi^2} M^4 m_Q^3 f_{\cal P}  
\Big[ I_2 - 
m_Q^2 I_3 \Big] \phi_\eta(u_0)        
+ {(1-\beta^2)\over 64 \sqrt{2} \pi^2} M^4 m_Q^2 \mu_{\cal P}
\Big\{ \Big(i_3({\cal T},1) - 2  i_3({\cal T},v)\Big) I_2 \nnb \\ 
\ek 2 m_Q^2 \Big[i_3({\cal T},1) - 2  i_3({\cal T},v) + (1-\widetilde{\mu}_{\cal P}^2)
\phi_\sigma(u_0) \Big] I_3 \Big\} \nnb \\
\ek
{(1-\beta)^2\over 128 \sqrt{2} \pi^2} M^2 m_{\cal P}^2 m_Q f_{\cal P}  
\Big\{ m_Q^2 \mathbb{A}(u_0) I_2 - 
2 \Big( i_2({\cal V}_\parallel,1) - 2 i_2({\cal V}_\perp,1) \Big)
I_1 \nnb \\
\ek 2 m_Q^2 \Big[ i_2({\cal A}_\parallel,1) -
2 \Big( i_2({\cal V}_\parallel,1) - i_2({\cal V}_\perp,1) +  i_2({\cal
A}_\parallel,v)\Big) \Big]I_2\Big\} \nnb \\
\ek  {(1-\beta^2)\over 96 \sqrt{2} \pi^2} M^2
\Big[ 3 m_{\cal P}^2 m_Q^2 \mu_{\cal P} \Big( 2 m_Q^2
I_3 - I_2 \Big)
\Big( i_2({\cal T},1) - 2  i_2({\cal T},v) \Big)  - 4 \dd f_{\cal P} \pi^2 
\phi_\eta(u_0) \Big] \nnb \\
\ar {(1-\beta^2)\over 384 \sqrt{2} M^6} m_{\cal P}^2 m_Q^4 m_0^2 f_{\cal P}
\dd \mathbb{A}(u_0) \nnb \\
\ek {1\over 2304 \sqrt{2} M^4} m_Q^2 m_0^2 \dd \Big\{
(1-\beta^2) m_{\cal P}^2 f_{\cal P}
\Big[ 5 \mathbb{A}(u_0) + 12 \Big(i_2({\cal A}_\parallel,1) +   
i_2({\cal V}_\parallel,1) - 2 i_2({\cal A}_\parallel,v) \Big) \Big] \nnb \\
\ek 8 m_Q \mu_{\cal P} (1-\widetilde{\mu}_{\cal P}^2)
[3 + \beta (2 + 3 \beta)] \phi_\sigma(u_0) \Big\} \nnb \\
\ek {1\over 864 \sqrt{2} M^2} m_Q \dd \Big\{9 (1-\beta^2) m_Q f_{\cal P}
\Big( m_{\cal P}^2 \mathbb{A}(u_0) + m_0^2 \phi_\eta(u_0) \Big) +
2 [5 + \beta (4 + 5 \beta)] m_0^2 \mu_{\cal P} (1-\widetilde{\mu}_{\cal
P}^2) \phi_\sigma(u_0) \Big\} \nnb \\
\ek {1\over 576 \sqrt{2}} \Big\{(1-\beta^2) f_{\cal P} \dd
\Big[ 6 m_{\cal P}^2 \mathbb{A}(u_0) - 12 m_{\cal P}^2 
\Big(i_2({\cal A}_\parallel,1) +
i_2({\cal V}_\parallel,1) - 2 i_2({\cal A}_\parallel,v) \Big) +
m_0^2 \phi_\eta(u_0) \Big] \nnb \\
\ar 8[3 + \beta (2 + 3 \beta)] m_Q \mu_{\cal P}
(1-\widetilde{\mu}_{\cal P}^2) \dd \phi_\sigma(u_0) \Big\}
\eea
where $I_n$ is defined as:
\bea
\label{nolabel}
I_n \es \int_{m_Q^2}^\infty ds 
{e^{m_Q^2/M^2 - s/M^2}\over s^n}~,\nnb
\eea
and other parameters and functions as well as the way of continuum subtraction are given in \cite{Rstp11}. 
To shorten the equation, we have ignored the light quark masses as well as terms containing gluon condensates in the above equation, but we 
take into account
 their contributions in numerical analysis.

\section{Numerical results}

In this section, we present the numerical results of the sum rules for strong
coupling constants of pseudoscalar mesons with sextet and antitriplet heavy
baryons, which are obtained in the previous section. The main input
parameters of LCSR are DA's for the pseudoscalar mesons which are given in
\cite{Rstp15,Rstp16,Rstp17,Rstp18}. The other input parameters entering to
the sum rules are $\qq = -(0.24 \pm 0.001)^3~GeV^3$, $m_0^2 = (0.8 \pm
0.2)~GeV^2$ \cite{Rstp22}, $f_\pi = 0.131~GeV$, $f_K = 0.16~GeV$ and 
$f_\eta = 0.13~GeV$ \cite{Rstp15}. 

The sum rules for the SSP, SAP and AAP coupling constants have three
auxiliary parameters: Borel mass parameter $M^2$, continuum threshold $s_0$
and the arbitrary parameter $\beta$ which exists in the expression in the
expression for the interpolating currents. Obviously, the result for any
measurable physical quantity, being coupling constant in the present case,
should be independent on them. Therefore, our primary goal is to find such
regions of these parameters, where coupling constants exhibits no
dependence.

The upper limit of $M^2$ is determined by requiring that the continuum and
higher states contributions should be small compared to the total dispersion
integral. The lower limit can be obtained from the condition that the
condensate terms with highest dimensions contributes smaller compared to the
sum of all terms.
These two conditions leads to the  working region,  $15~GeV^2 \le M^2 \le 30~GeV^2$ for the bottom baryons and
$4~GeV^2 \le M^2 \le 12~GeV^2$ for the charmed ones.
The continuum threshold is not totally arbitrary but it depends on the energy of the first excited state with the same quantum numbers as the interpolating current.
We choose it in  the domain between $s_0=(m_B + 0.5)^2~GeV^2$
and $s_0=(m_B + 1)^2~GeV^2$. As an example, let us consider the
$\Xi_b^{'0} \rar \Xi_b^{'0} \pi^0$ transition. In Fig. (2), the dependence of
the strong coupling constant for the $\Xi_b^{'0} \rar \Xi_b^{'0} \pi^0$
transition on $M^2$ is considered at different fixed values of $\beta$ and
a fixed value of $s_0$. We observe from this figure that the coupling
constant has a good stability in the ``working region" of $M^2$.  In Fig. (3), we present the dependence of the
strong coupling constant for the $\Xi_b^{'0} \rar \Xi_b^{'0} \pi^0$
transition on $\cos\theta$ at several fixed values of $s_0$
and at $M^2= 22.5~GeV^2$, where the angle $\theta$ is determined from
$\beta=\tan\theta$. From this figure, we see that the dependence of the coupling constant on  $s_0$ diminishes when the  higher values of the continuum threshold 
 are chosen from the considered working region. From this figure, we also observe that the strong coupling constant for the $\Xi_b^{'0} \rar \Xi_b^{'0} \pi^0$ decay
becomes very large near the end points ($\cos\theta=\pm1$) and have zeros at some finite values of the $\cos\theta$. This behavior can be explained as follows.
 From Eq. (\ref{estp20}) we see that the coupling constant is proportional to $ {1 \over \lambda_1^{(i)} \lambda_2^{(i)}} \Pi_1^{(i)}$. In general, zero's of the nominator and 
denominator does not coincide since the OPE is truncated. In other words, calculations are not exact. For this reason, these points and any region between them are not 
reliable regions for determination of physical quantities and suitable regions for $\cos\theta$ should be far from these regions.
It follows from Fig. (3) that in the region, $-0.5 \le
\cos\theta \le + 0.3$, the coupling constant seems to be insensitive to the
variation of $\cos\theta$. Here, we should also stress that our numerical results lead to the  working region, $-0.6 \le
\cos\theta \le + 0.5$ common for masses of all heavy spin 1/2 baryons, which includes the working region of $\cos\theta$ for  the coupling constant.  
This region lie also inside the more wide interval of $\cos\theta$ obtained from analysis of the masses of the non strange heavy baryons 
 in \cite{E.Bagan,E.Bagan2,E.Bagan3}.
In general, the working region
 of $\cos\theta$ for masses and coupling constants can be different, but in some cases as occur in our problem these regions coincide.

Similar analysis for the strong coupling constants of the light pseudoscalar
mesons with sextet and antitriplet heavy baryons are performed and the
results are presented in Tables (2), (3) and (4). In these Tables we also
present the predictions for the coupling constants coming from the Ioffe
currents when $\beta=-1$.
The errors in the values of the coupling constants presented in the
Tables (2), (3) and (4) include uncertainties coming from the variations of the
$s_0$, $\beta$ and $M^2$ as well as those coming from the other input parameters.


\begin{table}[t]

\renewcommand{\arraystretch}{1.3}
\addtolength{\arraycolsep}{-0.5pt}
\small
$$
\begin{array}{|l|r@{\pm}l|r@{\pm}l||l|r@{\pm}l|r@{\pm}l|}
\hline \hline  
 \multirow{2}{*}{$g^{\mbox{\small{\,channel}}}$}        &\multicolumn{4}{c||}{\mbox{Bottom Baryons}}   &  
 \multirow{2}{*}{$g^{\mbox{\small{\,channel}}}$}        &\multicolumn{4}{c|}{\mbox{Charmed Baryons}} \\
	                                                &   \multicolumn{2}{c}{\mbox{~General current~}}        & 
	                                                    \multicolumn{2}{c||}{\mbox{~Ioffe current~}}       & &
                                                                                                                   \multicolumn{2}{|c}{\mbox{~General current~}}  & 
                                                                                                                   \multicolumn{2}{c|}{\mbox{~Ioffe current~}}       \\ \hline
 g^{\Xi_b^{'0}    \rar \Xi_b^{'0}    \pi^0}     &~~~~~~~9.0&3.0    &~~~~~7.3&2.6 &
 g^{\Xi_c^{'+}    \rar \Xi_c^{'+}    \pi^0}     &~~~~~~ 4.0&1.4    &~~~~ 3.0&1.1   \\ 
 g^{\Sigma_b^0    \rar \Sigma_b^-    \pi^+}     &      17.0&6.1    &   13.0&4.5  &
 g^{\Sigma_c^+    \rar \Sigma_c^0    \pi^+}     &       8.0&2.8    &    4.1&1.5   \\
 g^{\Xi_b^{'0}    \rar \Sigma_b^+     K^-}      &      19.0&6.7    &   10.0&3.6  &
 g^{\Xi_c^{'+}    \rar \Sigma_c^{++}  K^-}      &       9.0&3.4    &    3.0&1.0   \\
 g^{\Omega_b^-    \rar \Xi_b^{'0}     K^-}      &      21.0&6.8    &   12.3&4.4  &
 g^{\Omega_c^0    \rar \Xi_c^{'+}     K^-}      &       9.0&3.4    &    5.6&1.9  \\
 g^{\Sigma_b^+    \rar \Sigma_b^+    \eta_1}    &      12.5&4.4    &    8.7&3.1  &
 g^{\Sigma_c^{++} \rar \Sigma_c^{++} \eta_1}    &       6.0&2.2    &    2.8&1.0   \\
 g^{\Xi_b^{'0}    \rar \Xi_b^{'0}    \eta_1}    &       5.3&1.9    &    3.6&1.3  &
 g^{\Xi_c^{'+}    \rar \Xi_c^{'+}    \eta_1}    &       2.6&0.9    &    0.7&0.2   \\
 g^{\Omega_b^-    \rar \Omega_b^-    \eta_1}    &      26.0&7.4    &   20.0&5.5  &
 g^{\Omega_c^0    \rar \Omega_c^0    \eta_1}    &      11.0&3.8    &    9.3&3.4   \\
 \hline \hline
\end{array}
$$

\caption{The values of the strong coupling constants $g$ for the transitions
among the sextet and sextet heavy baryons with pseudoscalar mesons.}

\renewcommand{\arraystretch}{1}
\addtolength{\arraycolsep}{-1.0pt}

\end{table}


\begin{table}[h]

\renewcommand{\arraystretch}{1.3}
\addtolength{\arraycolsep}{-0.5pt}
\small
$$
\begin{array}{|l|r@{\pm}l|r@{\pm}l||l|r@{\pm}l|r@{\pm}l|}
\hline \hline  
 \multirow{2}{*}{$g^{\mbox{\small{\,channel}}}$}        &\multicolumn{4}{c||}{\mbox{Bottom Baryons}}   &  
 \multirow{2}{*}{$g^{\mbox{\small{\,channel}}}$}        &\multicolumn{4}{c|}{\mbox{Charmed Baryons}} \\
	                                                &   \multicolumn{2}{c}{\mbox{~General current~}}        & 
	                                                    \multicolumn{2}{c||}{\mbox{~Ioffe current~}}       & &
                                                                                                                   \multicolumn{2}{|c}{\mbox{~General current~}}  & 
                                                                                                                   \multicolumn{2}{c|}{\mbox{~Ioffe current~}}       \\ \hline
 g^{\Xi_b^{'0}    \rar \Xi_b^0       \pi^0}     &~~~~~~~7.5&2.6    &~~~~~6.1&2.2  &
 g^{\Xi_c^{'+}    \rar \Xi_c^+       \pi^0}     &~~~~~~ 3.1&1.1    &~~~~ 2.0&0.7   \\
 g^{\Sigma_b^-    \rar \Lambda_b^0   \pi^-}     &      15.0&4.9    &   11.5&3.9  &
 g^{\Sigma_c^0    \rar \Lambda_c^+   \pi^-}     &       6.5&2.4    &    5.6&1.8  \\
 g^{\Sigma_b^0    \rar \Xi_b^0       \bar{K}^0} &      11.5&3.9    &    8.9&3.1  &
 g^{\Sigma_c^+    \rar \Xi_c^+       \bar{K}^0} &       5.0&1.7    &    3.7&1.3   \\
 g^{\Omega_b^-    \rar \Xi_b^-       \bar{K}^0} &      17.0&4.5    &   13.5&4.8  &
 g^{\Omega_c^0    \rar \Xi_c^0       \bar{K}^0} &       6.5&2.3    &    3.0&1.1   \\
 g^{\Xi_b^{'0}    \rar \Xi_b^-        K^+}      &      12.0&4.3    &    9.8&3.5  &
 g^{\Xi_c^{'+}    \rar \Xi_c^0        K^+}      &       4.5&1.6    &    2.1&0.8   \\
 g^{\Xi_b^{'0}    \rar \Xi_b^0       \eta_1}    &      16.0&5.6    &    12.0&4.3  &
 g^{\Xi_c^{'+}    \rar \Xi_c^+       \eta_1}    &       6.7&2.4    &    4.3&1.5   \\
 \hline \hline
\end{array}
$$

\caption{The values of the strong coupling constants $g$ for the transitions
among the sextet and antitriplet heavy baryons with pseudoscalar mesons.}

\renewcommand{\arraystretch}{1}
\addtolength{\arraycolsep}{-1.0pt}

\end{table}

\begin{table}[h]

\renewcommand{\arraystretch}{1.3}
\addtolength{\arraycolsep}{-0.5pt}
\small
$$
\begin{array}{|l|r@{\pm}l|r@{\pm}l||l|r@{\pm}l|r@{\pm}l|}
\hline \hline  
 \multirow{2}{*}{$g^{\mbox{\small{\,channel}}}$}        &\multicolumn{4}{c||}{\mbox{Bottom Baryons}}   &  
 \multirow{2}{*}{$g^{\mbox{\small{\,channel}}}$}        &\multicolumn{4}{c|}{\mbox{Charmed Baryons}} \\
	                                                &   \multicolumn{2}{c}{\mbox{~General current~}}        & 
	                                                    \multicolumn{2}{c||}{\mbox{~Ioffe current~}}       & &
                                                                                                                   \multicolumn{2}{|c}{\mbox{~General current~}}  & 
                                                                                                                   \multicolumn{2}{c|}{\mbox{~Ioffe current~}}       \\ \hline
 g^{\Xi_b^0       \rar \Xi_b^0       \pi^0}     &~~~~~~~1.0&0.3    &~~~~~4.0&1.4  &
 g^{\Xi_c^+       \rar \Xi_c^+       \pi^0}     &~~~~~~ 0.70&0.22  &~~~~ 2.7&0.9   \\
 g^{\Xi_b^-       \rar \Lambda_b^0    K^-}      &       1.5&0.5    &    5.2&1.8  &
 g^{\Xi_c^0       \rar \Lambda_c^+    K^-}      &       0.9&0.3    &    2.2&0.7   \\
 g^{\Xi_b^0       \rar \Xi_b^0       \eta_1}    &       0.6&0.2    &    2.9&1.0  &
 g^{\Xi_c^+       \rar \Xi_c^+       \eta_1}    &      0.07&0.02   &   0.26&0.08  \\
 g^{\Lambda_b^0   \rar \Lambda_b^0   \eta_1}    &       1.0&0.3    &    4.0&1.1  &
 g^{\Lambda_c^+   \rar \Lambda_c^+   \eta_1}    &      0.75&0.24   &    1.9&0.66  \\
 \hline \hline
\end{array}
$$

\caption{The values of the strong coupling constants $g$ for the transitions
among the antitriplet and antitriplet heavy baryons with pseudoscalar mesons.}

\renewcommand{\arraystretch}{1}
\addtolength{\arraycolsep}{-1.0pt}

\end{table}

\newpage

We see from these Tables that, there is substantial difference between the
predictions of the general current and the Ioffe current, especially for the
strong coupling constants of the antitriplet--antitriplet heavy baryons 
with pseoduscalar mesons, which can be explained as follows. As a result of 
the analysis of the dependence of the coupling constants on $\cos\theta$ 
we see that the value $\beta=-1$ belongs to the unstable region. Therefore, a
prediction at this point of $\beta$ is not reliable. 

Finally, we compare our results with those existing in literature. In
various works, the coupling constant $\Sigma_c \rar \Lambda_c \pi$ is 
estimated to be
\bea
\label{nolabel}
g^{\Sigma_c \rar \Lambda_c \pi} = \left\{ \begin{array}{l}
8.88,~\mbox{\rm \cite{Rstp23}~(relativistic three--quark model)}~,\\
6.82,~\mbox{\rm \cite{Rstp24}~(light--front quark model)},\\
10.8 \pm 2.2,~\mbox{\rm \cite{Rstp09}~(LCSR)}~,\\
6.5 \pm 2.4,~\mbox{\rm (our result)~(LCSR)}~.  
\end{array} \right. \nnb
\eea
We see that, within errors our result is close to the results of
\cite{Rstp09,Rstp23,Rstp24}. The coupling constant for the $\Xi_Q \Xi_Q \pi$ 
transition LCSR is estimated to have the values 
$g^{\Xi_c \rar \Xi_c \pi} = 1.0 \pm 0.5$ and $g^{\Xi_b \rar \Xi_b
\pi} = 1.6 \pm 0.4$, which are slightly larger compared to our
predictions. Finally, the coupling constant $g^{\Sigma_c \rar \Sigma_c
\pi}$ is calculated in \cite{Rstp09} and it is obtained that
$g^{\Sigma_c \rar \Sigma_c \pi} = -8.0 \pm 1.7$, which is in quite a
good agreement with our prediction.

In conclusion, the strong coupling constants of light pseudoscalar mesons with
sextet and antitriplet heavy baryons are studied within LCSR. It is shown
that, all coupling constants for the sextet--sextet, sextet--antitriplet and
antitriplet--antitriplet transitions are described by only one invariant
function in each class.

\section*{Acknowledgment}
The authors thank to A. Ozpineci for useful discussions.

\bAPP{A}{}

Here in this appendix, we present the expressions of the correlation functions
in terms of invariant function $\Pi_1^{(i)}$ involving 
$\pi$, $K$ and  $\eta$ mesons.

\begin{itemize}
\item Correlation functions describing pseudoscalar mesons with
sextet--sextet baryons.
\end{itemize}
\baeeq
\label{nolabel}
{1\over \sqrt{2}} \Pi^{\Sigma_b^+ \rar \Sigma_b^0     \pi^+ } \es 
\Pi^{\Xi_b^{'0} \rar \Sigma_b^0 \bar{K}^0 } =
\Pi^{\Sigma_b^0 \rar \Xi_b^{'0}      K^0  } = 
\Pi_1^{(1)}(d,u,b)~, \nnb \\
\Pi^{\Xi_b^{'0} \rar \Xi_b^{'-}     \pi^+ } \es
\Pi_1^{(1)}(d,s,b)~, \nnb \\
{1\over \sqrt{2}} \Pi^{\Sigma_b^- \rar \Sigma_b^0     \pi^- } \es
\Pi^{\Xi_b^{'-} \rar \Sigma_b^0      K^-  } =
\Pi^{\Sigma_b^0 \rar \Xi_b^{'-}      K^+  } = 
\Pi_1^{(1)}(u,d,b)~, \nnb \\
\Pi^{\Xi_b^{'-} \rar \Xi_b^{'0}     \pi^- } \es                          
\Pi_1^{(1)}(u,s,b)~, \nnb \\
{1\over \sqrt{2}} \Pi^{\Xi_b^{'0} \rar \Sigma_b^+      K^-  } \es
{1\over \sqrt{2}} \Pi^{\Sigma_b^+ \rar \Xi_b^{'0}      K^+  } = 
{\sqrt{6} \over 2} \Pi^{\Sigma_b^+ \rar \Sigma_b^+     \eta_1} =
\Pi_1^{(1)}(u,u,b)~, \nnb \\
{1\over \sqrt{2}} \Pi^{\Omega_b^- \rar \Xi_b^{'0}      K^-  } \es
{1\over \sqrt{2}} \Pi^{\Xi_b^{'0} \rar \Omega_b^-      K^+  } = 
{1\over \sqrt{2}} \Pi^{\Omega_b^- \rar \Xi_b^{'-} \bar{K}^0 } =
{1\over \sqrt{2}} \Pi^{\Xi_b^{'-} \rar \Omega_b^-      K^0  } =
- {\sqrt{6}\over 4} \Pi^{\Omega_b^- \rar \Omega_b^-     \eta_1}
\Pi_1^{(1)}(s,s,b)~, \nnb \\
{1\over \sqrt{2}} \Pi^{\Xi_b^{'-} \rar \Sigma_b^- \bar{K}^0 } \es
{1\over \sqrt{2}} \Pi^{\Sigma_b^- \rar \Xi_b^{'-}      K^0  } =
{\sqrt{6} \over 2} \Pi^{\Sigma_b^- \rar \Sigma_b^-     \eta_1} =
\Pi_1^{(1)}(d,d,b)~, \nnb \\
\Pi^{\Xi_b^{'0} \rar \Xi_b^{'0}     \eta_1} \es   {1\over \sqrt{6}} \Big[\Pi_1^{(1)}(u,s,b) - 2 \Pi_1^{(1)}(s,u,b)\Big]~, \nnb \\
\Pi^{\Xi_b^{'-} \rar \Xi_b^{'-}     \eta_1} \es   {1\over \sqrt{6}} \Big[\Pi_1^{(1)}(d,s,b) - 2 \Pi_1^{(1)}(s,d,b)\Big]~, \nnb \\
\eaeeq

\begin{itemize}
\item Correlation functions responsible for the transitions of the
sextet--antitriplet baryons.
\end{itemize}
\baeeq
\label{nolabel}
\sqrt{2} \Pi^{\Xi_b^{'0} \rar \Xi_b^0        \pi^0 } \es   
\Pi^{\Xi_b^{'0} \rar \Xi_b^-        \pi^+ } =
\Pi^{\Xi_b^{'-} \rar \Xi_b^0         K^-  } =
\Pi_1^{(2)}(u,s,b)~, \nnb \\ 
- \sqrt{2} \Pi^{\Xi_b^{'-} \rar \Xi_b^-        \pi^0 } \es 
\Pi^{\Xi_b^{'-} \rar \Xi_b^0        \pi^- } =
\Pi_1^{(2)}(d,s,b)~, \nnb \\ 
\Pi^{\Sigma_b^0 \rar \Lambda_b^0    \pi^0 } \es   {1\over \sqrt{2}} \Big[\Pi_1^{(2)}(u,d,b) + \Pi_1^{(2)}(d,u,b)\Big]~, \nnb \\ 
\Pi^{\Sigma_b^- \rar \Lambda_b^0    \pi^- } \es
- \Pi^{\Sigma_b^0 \rar \Xi_b^-         K^+  } =
\Pi_1^{(2)}(u,d,b)~, \nnb \\ 
- {1\over \sqrt{2}} \Pi^{\Sigma_b^+ \rar \Lambda_b^0    \pi^+ } \es         
- \Pi^{\Sigma_b^0 \rar \Xi_b^0    \bar{K}^0 } =
- \Pi^{\Xi_b^{'0} \rar \Lambda_b^0\bar{K}^0 } =
\Pi^{\Xi_b^{'0} \rar \Xi_b^-         K^+  } =
\Pi_1^{(2)}(d,u,b)~, \nnb \\ 
- {1\over \sqrt{2}} \Pi^{\Sigma_b^- \rar \Xi_b^-    \bar{K}^0 } \es
- {1\over \sqrt{2}} \Pi^{\Sigma_b^- \rar \Lambda_b^0     K^-  } =
\Pi_1^{(2)}(d,d,b)~, \nnb \\
{1\over \sqrt{2}} \Pi^{\Omega_b^- \rar \Xi_b^-    \bar{K}^0 } \es
{1\over \sqrt{2}} \Pi^{\Omega_b^- \rar \Xi_b^0         K^-  } =
\Pi_1^{(2)}(s,s,b)~, \nnb \\
\Pi^{\Sigma_b^+ \rar \Lambda_b^0     K^+  } \es -         \sqrt{2}       \Pi_1^{(2)}(u,u,b)~, \nnb \\  
\Pi^{\Xi_b^{'0} \rar \Xi_b^0        \eta_1} \es   {1\over \sqrt{6}} \Big[\Pi_1^{(2)}(u,s,b) + 2 \Pi_1^{(2)}(s,u,b)\Big]~, \nnb \\
\Pi^{\Xi_b^{'-} \rar \Xi_b^-        \eta_1} \es   {1\over \sqrt{6}} \Big[\Pi_1^{(2)}(d,s,b) + 2 \Pi_1^{(2)}(s,d,b)\Big]~, \nnb \\
\Pi^{\Sigma_b^0 \rar \Lambda_b^0    \eta_1} \es   {1\over \sqrt{6}} \Big[\Pi_1^{(2)}(u,d,b) -   \Pi_1^{(2)}(d,u,b)\Big]~. \nnb
\eaeeq 

\begin{itemize}
\item Correlation functions appearing in the
antitriplet--antitriplet pseudoscalar meson transitions.
\end{itemize}
\baeeq
\label{nolabel}
\sqrt{2} \Pi^{\Xi_b^0    \rar \Xi_b^0        \pi^0 } \es
\Pi^{\Xi_b^0    \rar \Xi_b^-        \pi^+ } =
\Pi_1^{(3)}(u,s,b)~, \nnb \\
- \sqrt{2} \Pi^{\Xi_b^-    \rar \Xi_b^-        \pi^0 } \es
\Pi^{\Xi_b^-    \rar \Xi_b^0        \pi^- } =
\Pi_1^{(3)}(d,s,b)~, \nnb \\ 
\Pi^{\Lambda_b^0\rar \Lambda_b^0    \pi^0 } \es   {1\over \sqrt{2}} \Big[\Pi_1^{(3)}(u,d,b) - \Pi_1^{(3)}(d,u,b)\Big]~, \nnb \\ 
\Pi^{\Xi_b^0    \rar \Lambda_b^0\bar{K}^0 } \es                          \Pi_1^{(3)}(u,u,b)~, \nnb \\  
\Pi^{\Xi_b^-    \rar \Lambda_b^0     K^-  } \es -                        \Pi_1^{(3)}(u,d,b)~, \nnb \\  
\Pi^{\Xi_b^0    \rar \Xi_b^0        \eta_1} \es   {1\over \sqrt{6}} \Big[\Pi_1^{(3)}(u,s,b) - 2 \Pi_1^{(3)}(s,u,b)\Big]~, \nnb \\
\Pi^{\Xi_b^-    \rar \Xi_b^-        \eta_1} \es   {1\over \sqrt{6}} \Big[\Pi_1^{(3)}(d,s,b) - 2 \Pi_1^{(3)}(s,d,b)\Big]~, \nnb \\
\Pi^{\Lambda_b^0\rar \Lambda_b^0    \eta_1} \es   {1\over \sqrt{6}} \Big[\Pi_1^{(3)}(d,u,b) +   \Pi_1^{(3)}(u,d,b)\Big]~. \nnb
\eaeeq 

The expressions for the charmed baryons can easily be obtained by making the
replacement $b \rar c$ and adding to charge of each baryon a positive unit
charge. 

\eAPP

\begin{figure}[t]
\begin{center}
\scalebox{0.7}{\includegraphics{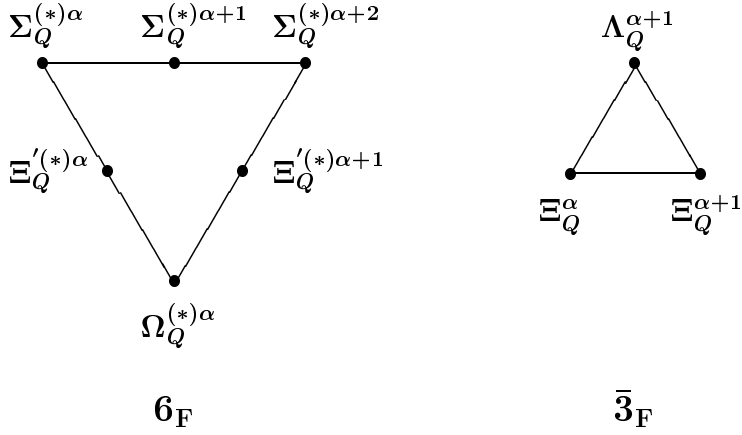}}
\end{center}
\caption{Sextet $(6_F)$ and antitriplet $(\bar{3}_F)$ representations 
of heavy baryons. Here $\alpha$, $\alpha +1$, $\alpha +2$ determine 
the charges of baryons $(\alpha=-1$ or $ 0)$, and $(\ast)$ denote 
$J^P={3\over 2}^+$ states.}
\end{figure}

\begin{figure}[b]
\begin{center}
\scalebox{0.8}{\includegraphics{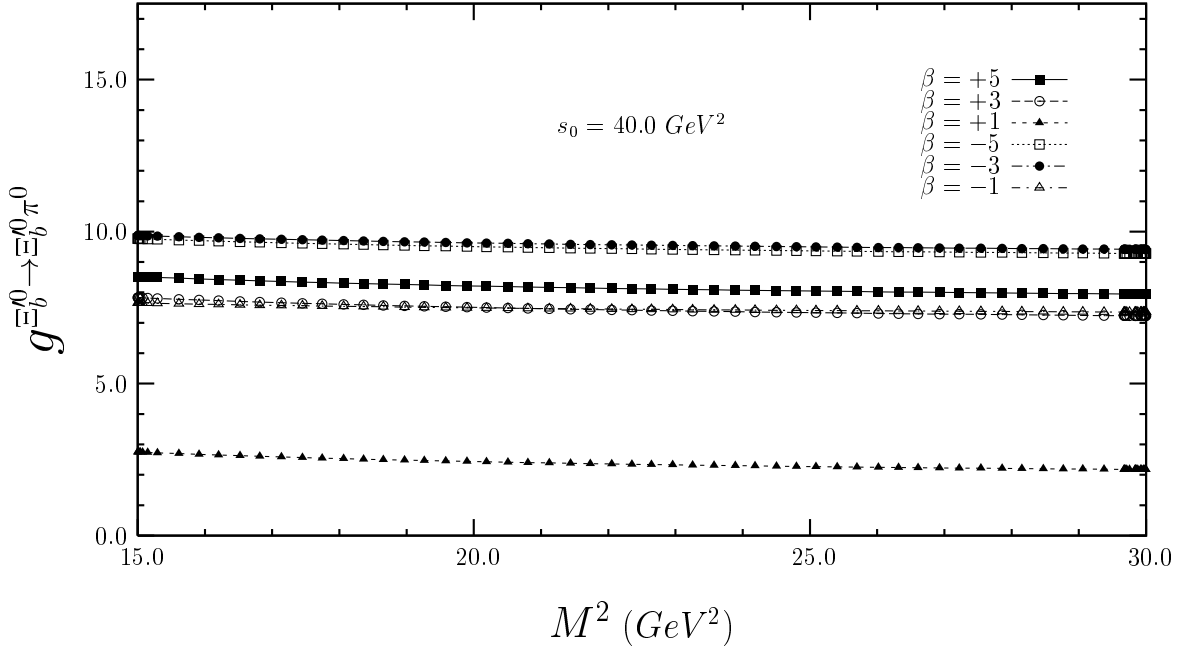}}
\end{center}
\caption{The dependence of the strong coupling constant for 
the $\Xi_b^{'0} \rar \Xi_b^{'0} \pi^0$ transition at several 
different fixed values of $\beta$, and at $s_0=40.0~GeV^2$.}
\end{figure}

\begin{figure}[t]
\begin{center}
\scalebox{0.8}{\includegraphics{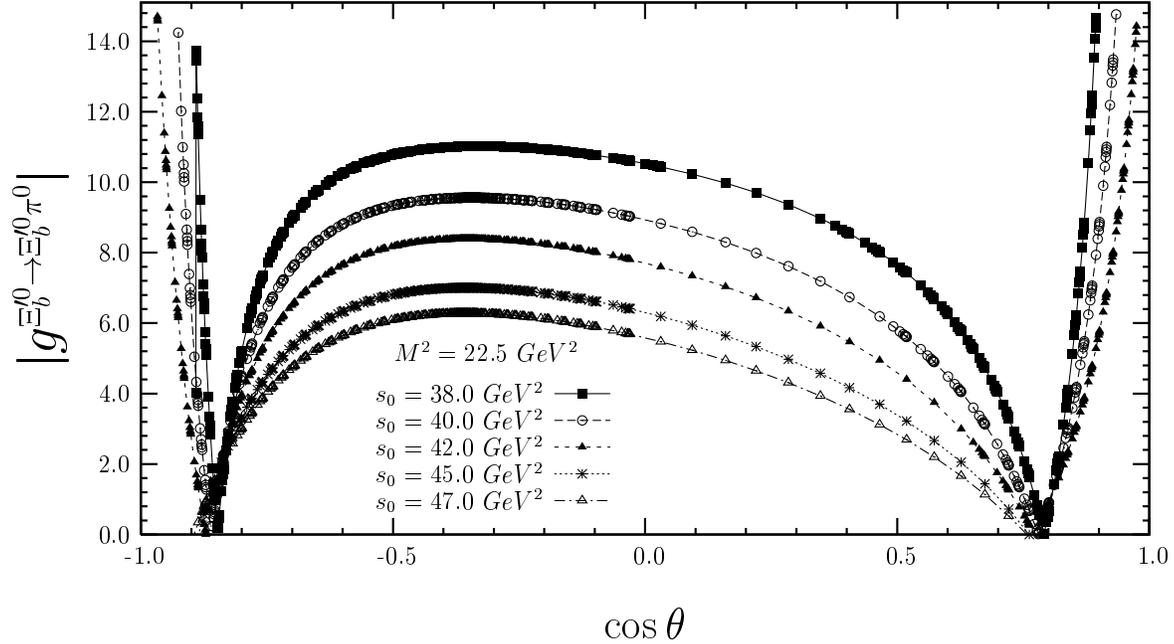}}
\end{center}
\caption{The dependence of the strong coupling constant for 
the $\Xi_b^{'0} \rar \Xi_b^{'0} \pi^0$ transition on $\cos\theta$ 
at several different fixed values of $s_0$, and at $M^2= 22.5~GeV^2$.}
\end{figure}

\end{document}